\documentclass[prl,aps,twocolumn,showpacs,superscriptaddress]{revtex4-1}

\usepackage{graphicx}
\usepackage{dcolumn}
\usepackage{bm}
\usepackage{float}
\begin{document}


\title{%
  Persistence and eventual demise of oxygen molecules at terapascal
  pressures}

\author{Jian Sun}
\email[]{jian.sun@theochem.rub.de}
\affiliation{Lehrstuhl f\"ur Theoretische Chemie, Ruhr-Universit\"at
  Bochum, 44780 Bochum, Germany}

\author{Miguel Martinez-Canales}
\affiliation{Department of Physics and Astronomy,
University College London, Gower Street, London WC1E 6BT, United Kingdom}

\author{Dennis D.\ Klug}
\affiliation{Steacie Institute for Molecular
  Sciences, National Research Council of Canada, Ottawa, K1A 0R6,
  Canada}

\author{Chris J.\ Pickard}
\affiliation{Department of Physics and Astronomy,
University College London, Gower Street, London WC1E 6BT, United Kingdom}

\author{Richard J.\ Needs}
\affiliation{Theory of Condensed Matter Group,
Cavendish Laboratory, J J Thomson Avenue, Cambridge CB3 0HE,
  United Kingdom}



\date{\today}

\begin{abstract}
  Computational searches for structures of solid oxygen under
  pressures in the multi TPa range have been carried out using
  density-functional-theory methods.  We find that molecular oxygen
  persists to about 1.9 TPa at which it transforms into a
  semiconducting square spiral-like polymeric structure ($I4_1/acd$)
  with a band gap of $\sim$3.0 eV.  Solid oxygen forms a metallic
  zig-zag chain-like structure ($Cmcm$) at about 3.0 TPa, but the chains in
  each layer gradually merge as the pressure is increased and a
  structure of $Fmmm$ symmetry forms at about 9.5 TPa in which each
  atom has four nearest neighbors.  The superconducting properties of
  molecular oxygen do not vary much with compression, although the
  structure becomes more symmetric.  The electronic properties of
  oxygen have a complex evolution with pressure, swapping between
  insulating, semiconducting and metallic.
\end{abstract}




\maketitle




%

%
%

Among the simple molecules studied at high pressures, oxygen has
attracted particular attention due to its fundamental importance and
intriguing properties.
\cite{Shimizu1998,Goncharenko2005,Lundegaard2006,Fujihisa2006,Weck2009}
For example, oxygen is the only known elemental molecule that exhibits
magnetism, which adds substantial complexity to its phase diagram.
When liquid oxygen is cooled at ambient pressure it undergoes a
sequence of transitions to the $\gamma$, $\beta$ and $\alpha$ solid
phases at 54.39, 43.76 and 23.88 K, respectively \cite{Barrett1967}.
Upon increase of pressure the monoclinic $\alpha$-O$_2$ ($C2/m$) phase
transforms into an orthorhombic $\delta$-O$_2$ ($Fmmm$) phase at about
3 GPa and to $\epsilon$-O$_2$ at about 10 GPa.  The structure of
$\epsilon$-O$_2$ has only recently been solved by x-ray diffraction
(XRD) studies of single crystal \cite{Lundegaard2006} and powder
\cite{Fujihisa2006} samples, although its vibrational spectrum was
reported more than 20 years earlier \cite{Nicol1979}.  The unit cell
of $\epsilon$-O$_2$ has $C2/m$ symmetry and contains four O$_2$
molecules forming O$_8$ units.  Both $\alpha$ and $\delta$-O$_2$ are
antiferromagnetic and the magnetic collapse at the $\delta$-$\epsilon$
transition was predicted using molecular dynamics simulations
\cite{Serra1998}.  This breakdown of the long-range antiferromagnetic
order at about 8 GPa was recently observed in a neutron scattering
experiment \cite{Goncharenko2005}.  Density functional theory (DFT)
studies have found another chain-like structure to be energetically
slightly more favorable than the O$_8$ structure \cite{Neaton2002,
  Ma2007}, although it has not been observed in experiments.

Recent experiments\cite{Weck2009} have shown that insulating $\epsilon$-O$_2$ remains
stable up to about 96 GPa before undergoing a continuous displacive
and isosymmetric transition to the $\zeta$ phase, in agreement with
earlier predictions \cite{Ma2007}.  The metallic $\zeta$ phase
\cite{Desgreniers1990,Akahama1995} has $C2/m$ symmetry
\cite{Goncharenko2005} and superconducts at temperatures below 0.6 K
\cite{Shimizu1998}.  Many-body perturbation theory $GW$ calculations
\cite{Kim2008a, Tse2008} have suggested that the metal-insulator
transition occurs at lower pressures than the measured
$\epsilon$-$\zeta$ transition pressure of 96 GPa.  The above
information, however, covers only a small part of the phase diagram
and rather little is known about pure oxygen at pressures above 100
GPa.

A previous DFT study reported that oxygen molecules persist to at
least 250 GPa \cite{Ma2007}. It is interesting to speculate about the
highest pressure to which oxygen molecules can survive, and whether
oxygen forms polymeric materials as found in N$_2$ \cite{Eremets2004,
  Pickard2009-nitrogen,Ma2009-nitrogen}, CO \cite{Lipp2005,
  Sun2011-CO}, and CO$_2$ \cite{Iota2007, Sun2009-CO2}, and predicted
in H$_2$ \cite{Pickard2007-hydrogen,McMahon2011}.  Materials under
terapascal pressures are of great interest in planetary science, for
example, the pressure at the center of Jupiter is estimated to be
about 7 TPa \cite{Jeanloz2007}.  Recent progress in dynamical shock
wave \cite{Jeanloz2007,Knudson2008,Eggert2010} and ramped compression
experiments \cite{Hawreliak2007, Bradley2009} has demonstrated that
the terapascal pressure regime is becoming much more accessible.

The use of DFT computations combined with searching methods has
provided a new route for predicting the structures and energetics of
high-pressure phases \cite{Sun2009-carbon,Pickard2010-Al,McMahon2011}.
A very recent study of phase transitions, melting, and chemical
reactivity in CO$_2$ found it to dissociate into carbon and oxygen
above 33 GPa and 1720 K \cite{Litasov2011}, which adds further
motivation for studying pure oxygen at high pressures.  In this work
we focus on oxygen at ultra-high pressures up to the terapascal
regime, finding very surprising behavior.

We have used the \textit{ab initio} random structure searching (AIRSS)
method \cite{Pickard2007-hydrogen,Pickard2009-nitrogen,Pickard2010-Al,
  Pickard2011-review} and DFT calculations to identify low-enthalpy
structures of oxygen in the multi TPa range.  We used the
Perdew-Burke-Ernzerhof (PBE) \cite{PBE} generalized gradient
approximation (GGA) exchange-correlation density functional.  The
searches were performed using the \textsc{castep} plane-wave DFT code
\cite{CASTEP} and ultrasoft pseudopotentials.  Searches were performed
at selected combinations of 0.5, 1, 2, 3, 4, 5 and 8 TPa, with 6, 8, 9, 10,
12 and 16 oxygen atoms per cell.  A total of about 14000 relaxed
structures were generated in the searches.  The enthalpy-pressure
relations were then recalculated using very hard projector augmented
wave (PAW) pseudopotentials and the \textsc{VASP} code \cite{vasp}
with a plane-wave basis set cutoff energy of 900 eV.  Phonon and
electron-phonon coupling calculations were performed using DFT
perturbation theory and the \textsc{abinit} code \cite{ABINIT},
norm-conserving pseudopotentials, the PBE functional, and an energy
cutoff of 1632 eV.  Further details of the calculations are given in
the supplementary material \cite{EPAPS}.

\begin{figure}[h]
\begin{center}
\includegraphics[width=8cm]{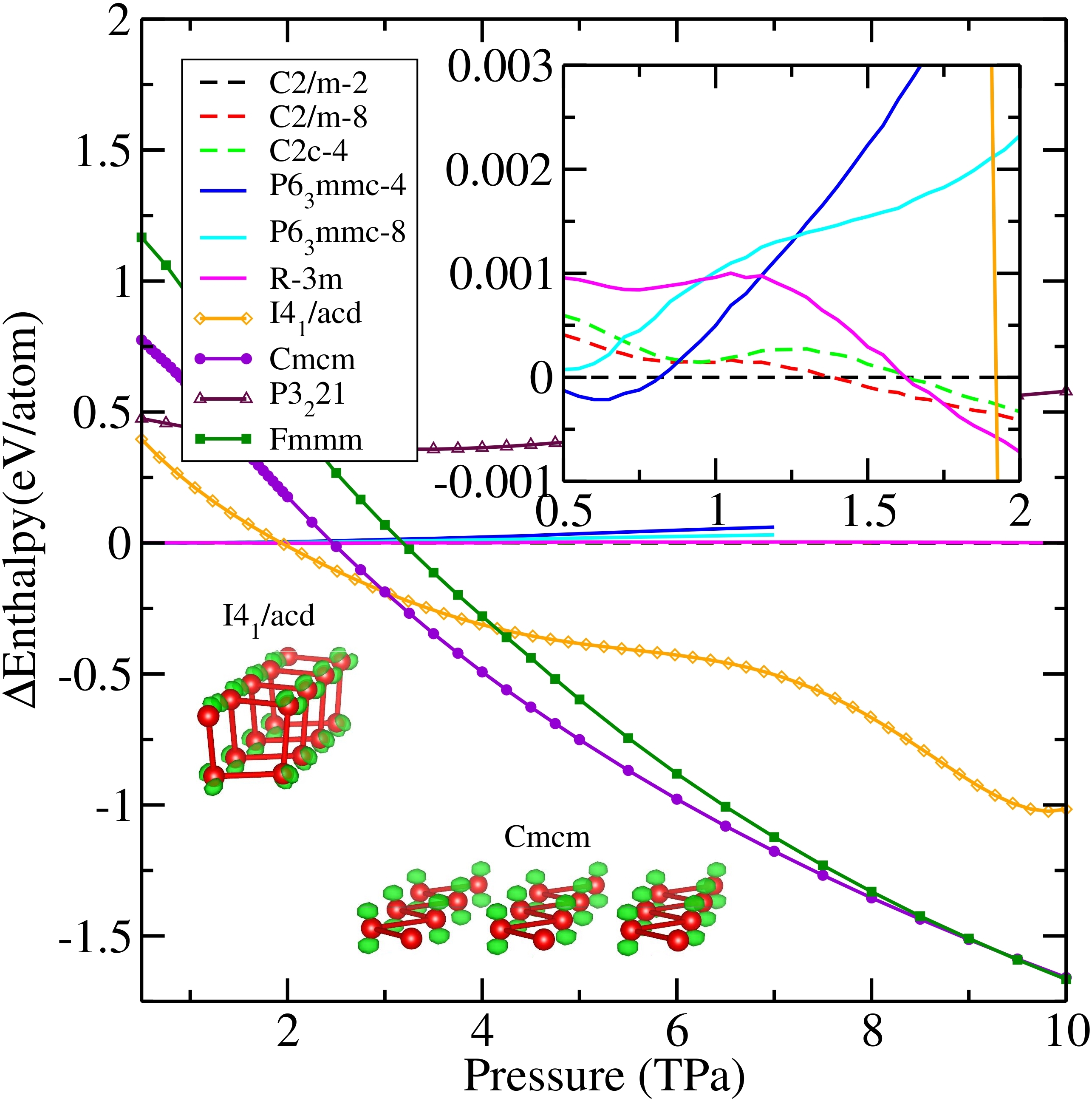}
\caption{%
  Enthalpy-pressure relations for solid oxygen. The upper right inset
  shows the enthalpies of the molecular phases, while the insets in
  the lower left corner show the square spiral structure of $I4_1/acd$
  and zig-zag chains of $Cmcm$, respectively.  The green bubbles
  represent the electron lone pairs.  }
\label{fig:eos}
\end{center}
\end{figure}

\begin{figure*}[htp]
\begin{center}
\includegraphics[width=0.3\textwidth]{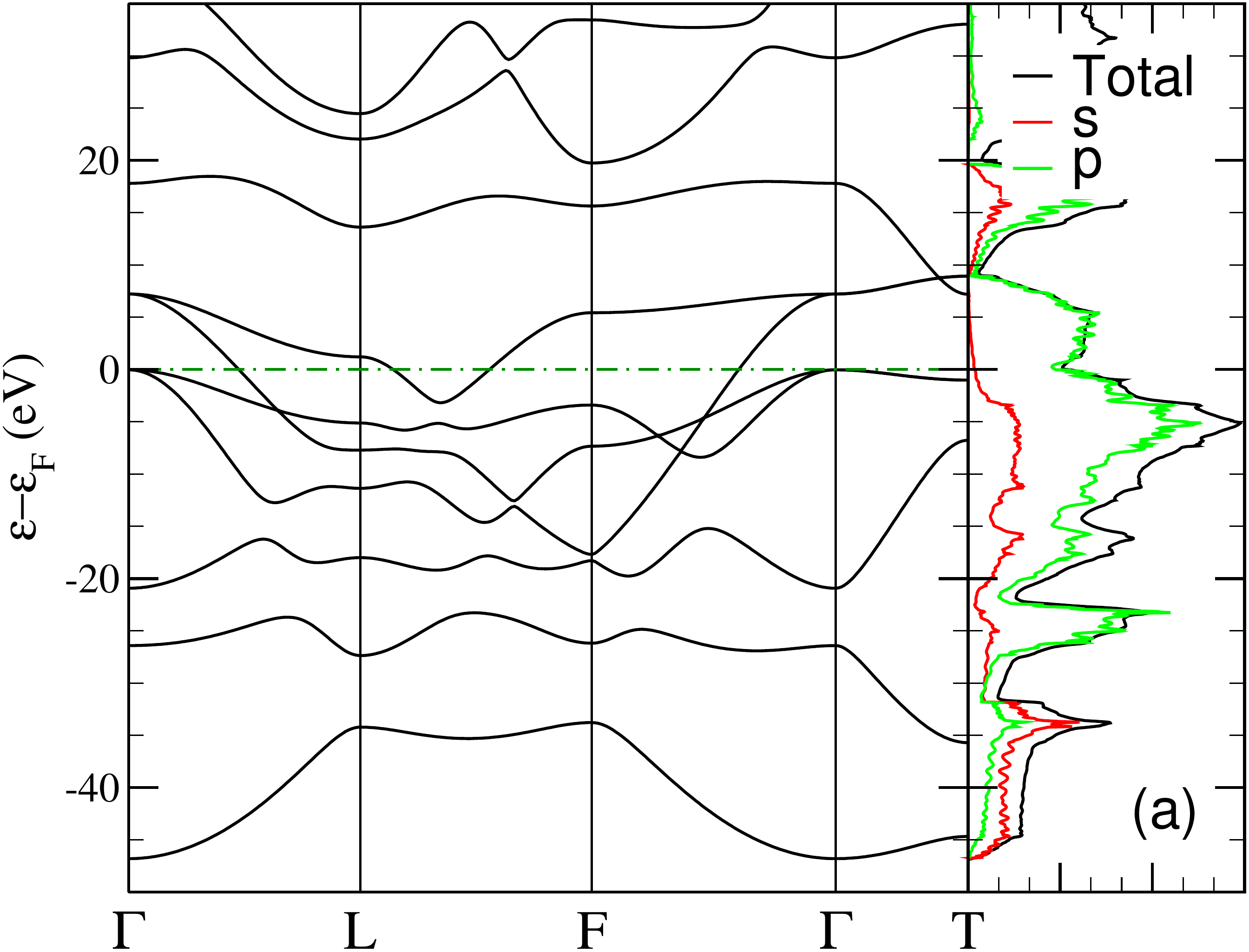}
\includegraphics[width=0.3\textwidth]{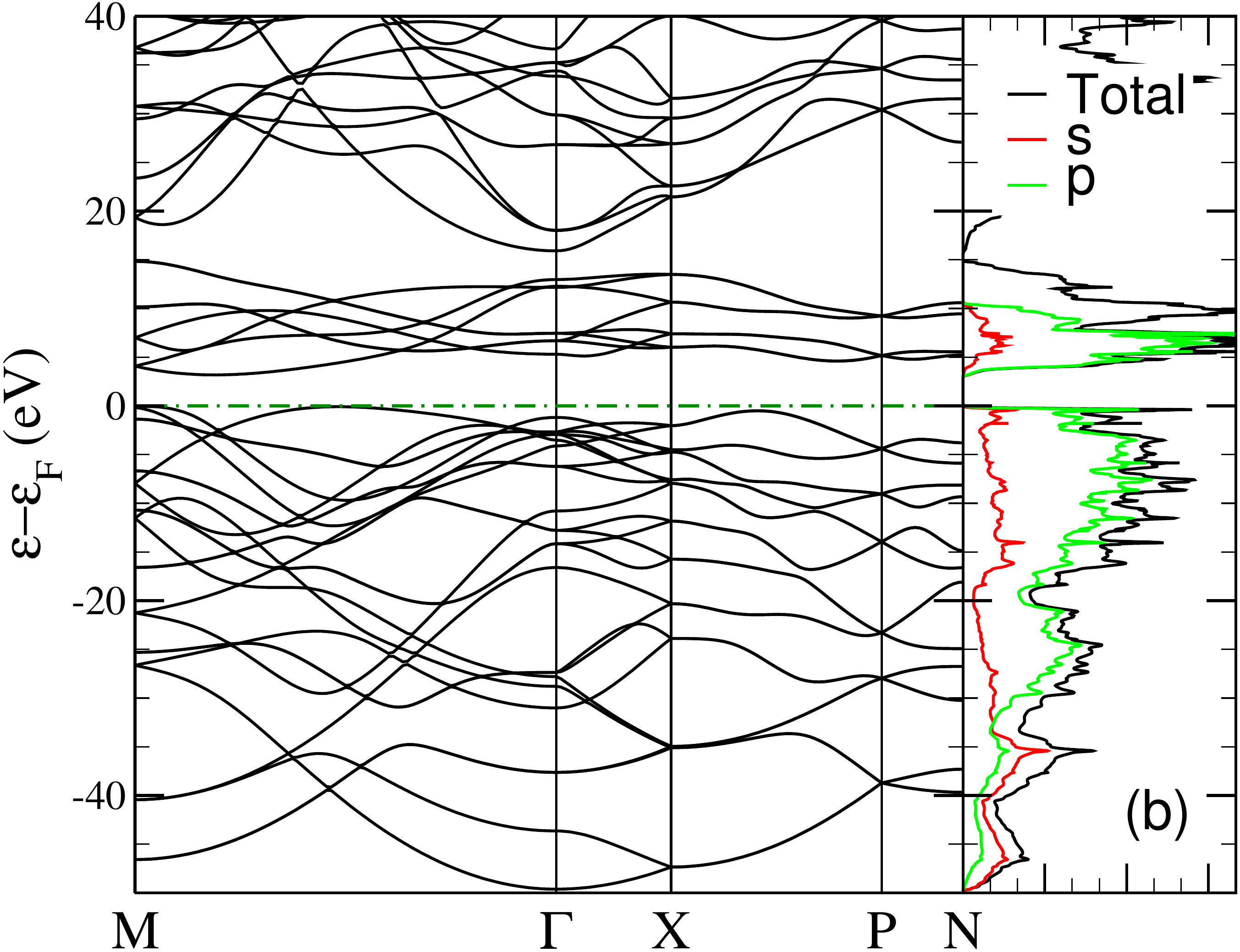}
\includegraphics[width=0.3\textwidth]{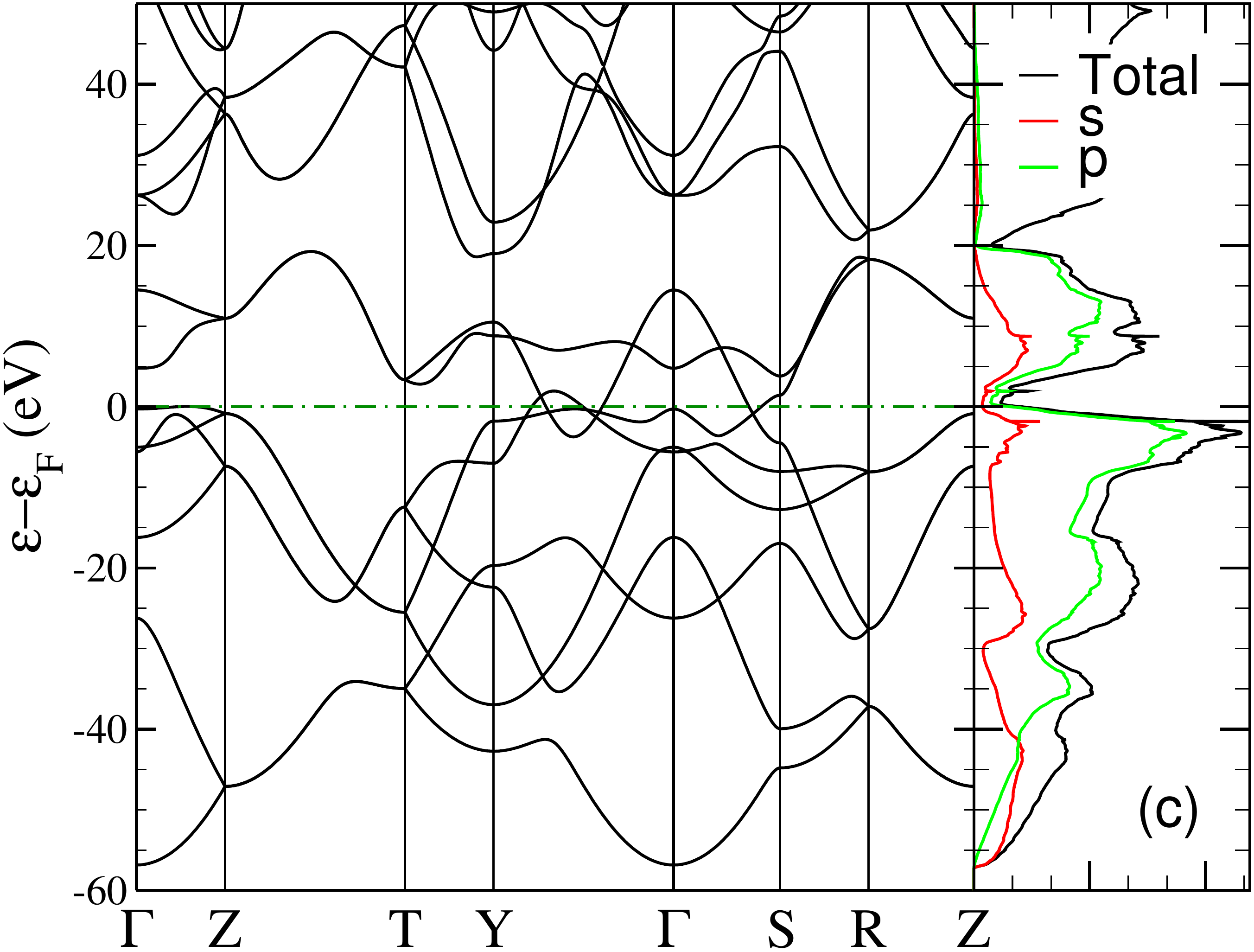}
\caption{%
  Electronic band structures and projected densities of states (PDoS).
  (a) $R\bar{3}m$ at 1.8 TPa, (b) $I4_1/acd$ at 2.0 TPa, and (c)
  $Cmcm$ at 3.5 TPa.  The zero of energy is at the Fermi level.  }
\label{fig:band}
\end{center}
\end{figure*}

\begin{figure}[h]
\begin{center}
\includegraphics[width=0.5\textwidth]{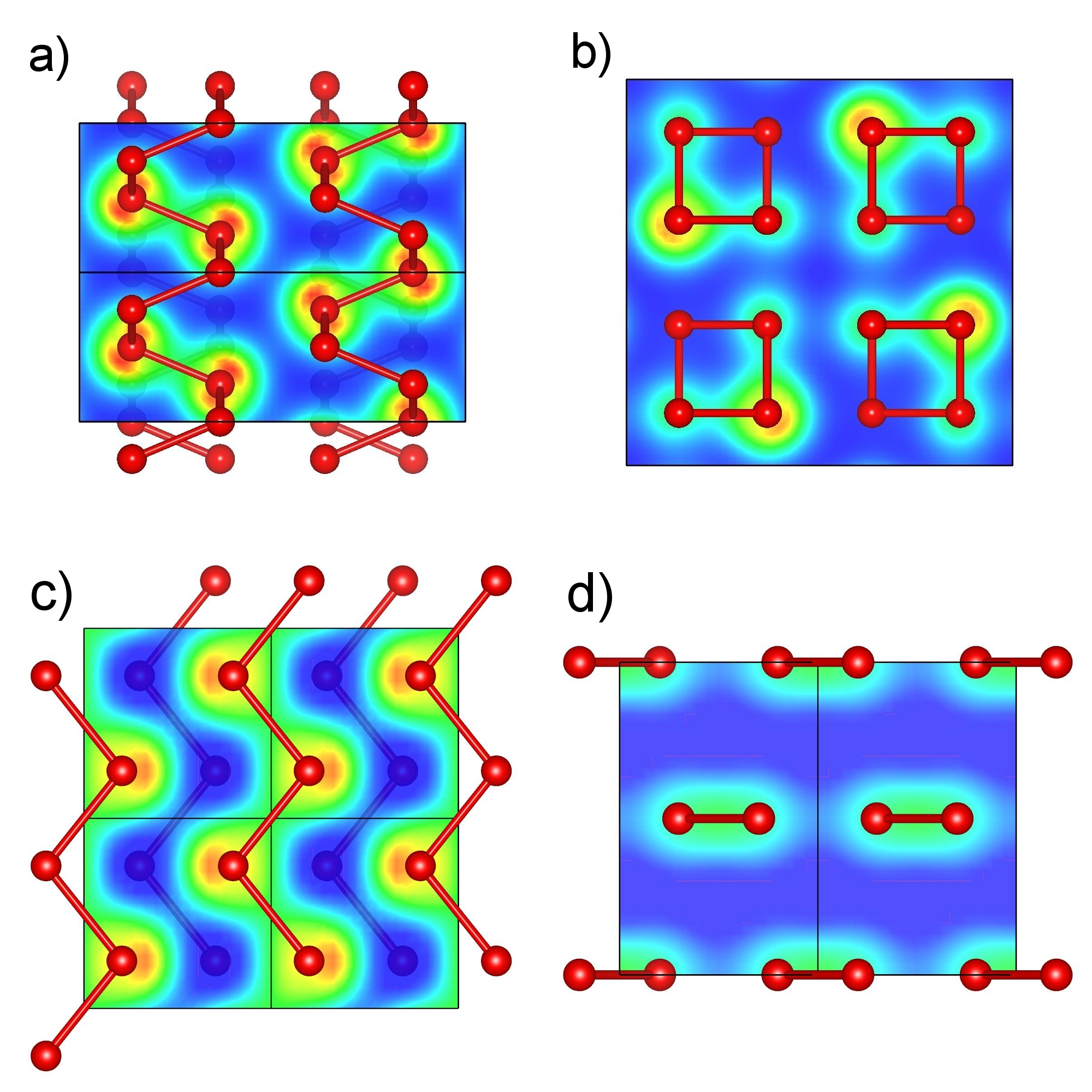}
\caption{%
  Electronic densities of polymeric oxygen.  (a) and (b) $I4_1/acd$ at
  2.0 TPa along the [100] and [001] directions, (c) and (d) $Cmcm$ at
  3.5 TPa along the [100] and [001] directions.  }
\label{fig:charge_density}
\end{center}
\end{figure}

The enthalpy-pressure relations for the most interesting structures
are plotted in Fig.\ \ref{fig:eos}.  We predict that at about 1.9 TPa
molecular oxygen should transform into a square spiral-like structure
belonging to space group $I4_1/acd$, as shown in the inset in the
lower left corner of Fig.\ \ref{fig:eos}.  Similar structures have
been found in other heavier group VI elements, such as sulfur and
selenium \cite{Degtyareva2005}.  The shortest O-O bond length in
$I4_1/acd$ at 2 TPa is about 1.15 \AA, which is longer than that of
the molecular phase at the same pressure (about 1.03 \AA).  The
shortest O-O distance between the chains is about 1.55 \AA, and the
distance between the axes of neighboring square spirals is half of the
lattice vector. The O-O-O angle in the helix is about $98.7^\circ$.

At about 3.0 TPa, the square-spiral phase transforms into a zig-zag
chain-like phase of $Cmcm$ symmetry, as shown in Fig.\ \ref{fig:eos}.
The O-O bond length within the chains is 1.10 \AA\ at 3.5 TPa, while
the inter-chain O-O separation in the plane is 1.33 \AA.  The O-O-O
angle within the chain is $102.8^\circ$.  With increasing pressure the
chains in each layer of $Cmcm$ gradually merge and at about 9.5 TPa a
structure of $Fmmm$ symmetry forms in which each atom has four nearest
neighbors at a separation of 1.05 \AA.

The low-pressure molecular phases are very close in enthalpy, as the
inset of Fig.\ \ref{fig:eos} shows.  Among these phases, $P6_3/mmc$-4,
$C2/m$-2, $C2/m$-8 and $R\bar{3}m$ become stable in turn on increasing
the pressure from 0.5 to 1.9 TPa.  More precise methods than DFT might
be required to clarify the structural sequence as the enthalpy
differences are so small.  Although the O=O bond in the O$_2$ molecule
is not as strong as the N$\equiv$N and C$\equiv$O triple bonds, O$_2$
polymerizes at a much higher pressure than N$_2$ ($\sim$110 GPa
\cite{Eremets2004}) and CO (experiments show that CO polymerizes at
about 5 GPa \cite{Lipp2005} while calculations predict that it could
even polymerize at ambient pressure \cite{Sun2011-CO}).  The fact that
the molecular phases persist to pressures as high as 1.9 TPa is
intriguing.  Electron counting arguments indicate that an oxygen atom
can form two covalent bonds, either a double bond as in the O$_2$
molecule or two single bonds, as would be expected in polymerized
oxygen.  The double bonds of the molecule are shorter and stronger
than the single bonds of polymerized forms, and the lone electron
pairs on the oxygen atoms result in bent bonds, as in the ozone O$_3$
molecule, and give rise to strongly repulsive interactions between the
molecules and between the polymeric chains.  The great reluctance of
oxygen to form more than two covalent bonds, the requirement that
these bonds be bent, and the lone-pair repulsion greatly limits the
forms of dense structures which may occur.  The square spirals of the
$I4_1/acd$ structure are not well packed and, while the zig-zag chains
of $Cmcm$ pack more efficiently, there is no bonding between chains so
that very dense structures are not formed.  These factors lead to the
persistence of molecular forms to high pressures, and the absence of
stable framework structures up to at least 10 TPa.

We studied other possible candidate structural types such as buckled
octagons of the type found in the S-I phase of sulfur ($Fddd$) and the
3-fold spiral-like $P3_221$ structure of S-II \cite{Degtyareva2005},
but these were found to be much less stable than the best structures
obtained in our searches.

The electronic band structures and projected electronic densities of
states (PDoS) of $R\bar{3}m$ at 1.8 TPa, $I4_1/acd$ at 2.0 TPa and
$Cmcm$ at 3.5 TPa are shown in Fig.\ \ref{fig:band}, decomposed into
$s$ and $p$ components.  As mentioned above, the enthalpies of the
molecular phases are very close, but the $R\bar{3}m$ structure is the
most likely molecular ground state prior to polymerization at 1.9 TPa.
Our DFT calculations suggest that the $R\bar{3}m$ structure is
metallic at 1.8 TPa and the electronic density of states at the Fermi
level mostly derives from the $p$ electrons.

Surprisingly, it turns out that the square spiral-like $I4_1/acd$
phase has a band gap of 3.0 eV at 2.0 TPa, as shown in Fig.\
\ref{fig:band}(b).  Considering that standard density functionals such
as PBE are well-known for underestimating band gaps, $I4_1/acd$ is
expected to be a wide-gap semiconductor.  As shown in the electron
density plots of Fig.\ \ref{fig:charge_density}(a) and (b), and also
in the charge density difference in the lower left inset of Fig.\
\ref{fig:eos}, the electrons are strongly localized.  Strong covalent
bonds are formed along the spiral chains and the projections of the
electron density on the [100] and [001] planes show that the spirals
are separated and not bonded to each other.

The calculated PDoS and band gap of the $Cmcm$ phase (Fig.\
\ref{fig:band}(c)) show that oxygen is metallic at 3.5 TPa.  The
density of states at the Fermi level derives approximately equally
from the $s$ and $p$ electrons and is smaller than that of the
$R\bar{3}m$ structure (Fig.\ \ref{fig:band}(a)).  This indicates that
the $Cmcm$ phase is a weak metal and supports the conclusion that the
lone pair electrons are not very extended.  The electron density plots
of Figs.\ \ref{fig:charge_density}(c) and (d) show that the zig-zag
chains are not bonded to one another and that the layers are well
separated.

\begin{figure}[h]
\begin{center}
\includegraphics[width=0.45\textwidth]{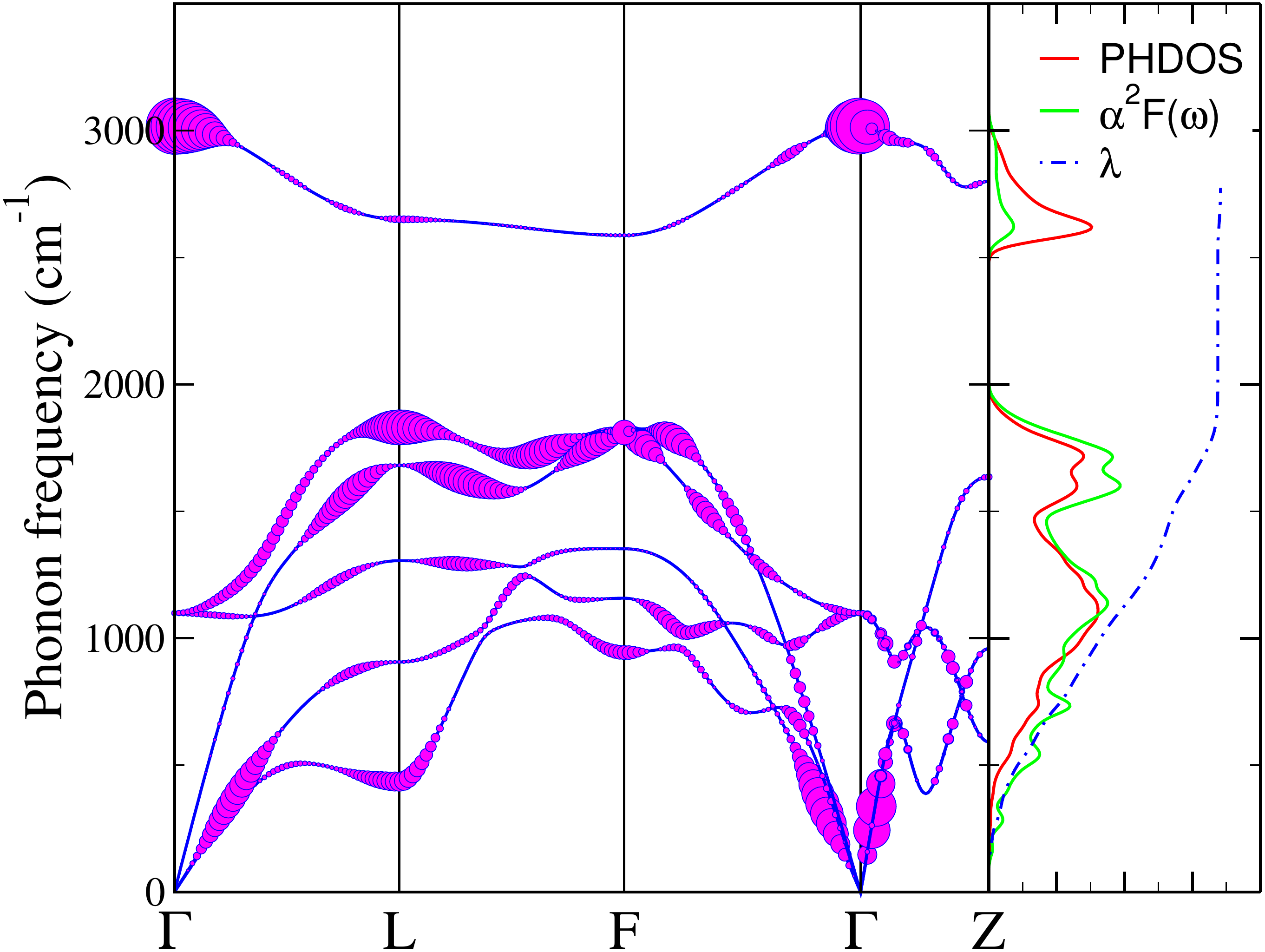}
\includegraphics[width=0.225\textwidth]{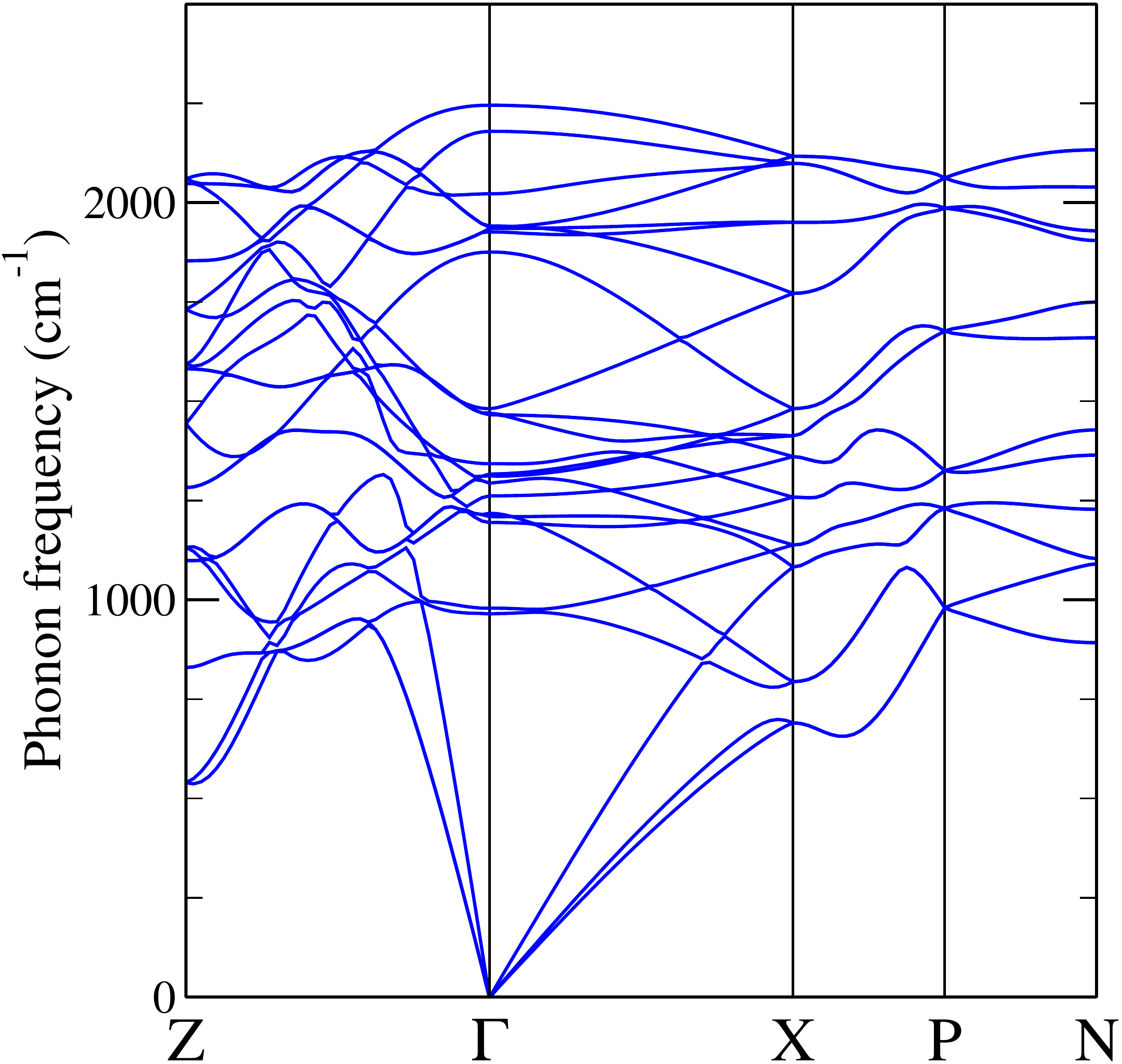}
\includegraphics[width=0.225\textwidth]{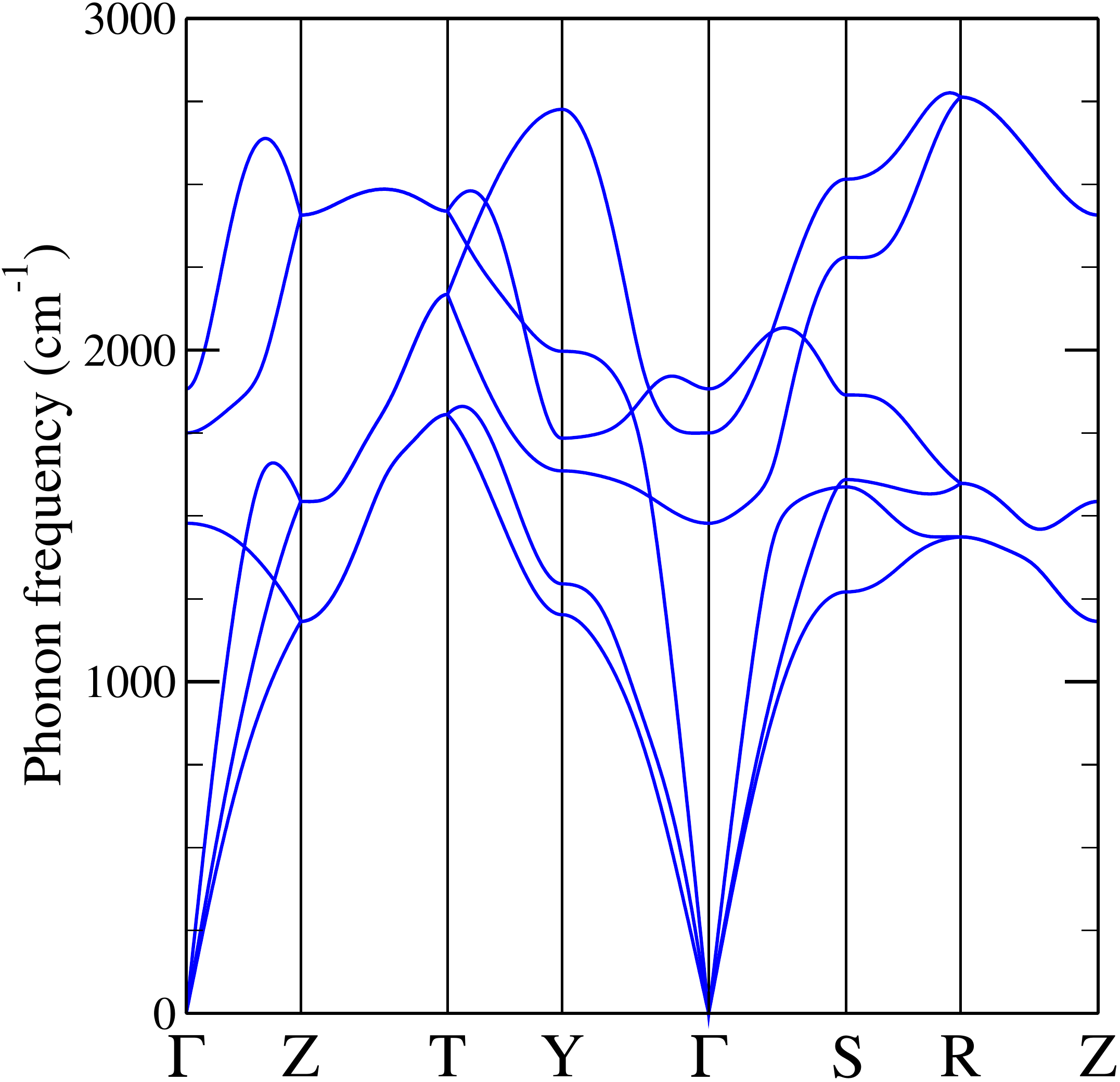}
\caption{%
  Phonon dispersion relations of $R\bar{3}m$ at 1.8 TPa (upper),
  $I4_1/acd$ at 2.0 TPa (lower left) and $Cmcm$ at 3.5 TPa (lower
  right).  The blue circles with magenta filling show the phonon
  linewidth at different wave vectors.  }
\label{fig:phonon}
\end{center}
\end{figure}

We investigated the dynamical stability of the $R\bar{3}m$, $I4_1/acd$
and $Cmcm$ structures by calculating their phonon dispersion
relations, as shown in Fig.\ \ref{fig:phonon}.  All of the structures
are mechanically stable within certain pressure ranges.  The phonons
of both the molecular $R\bar{3}m$ and $C2/m$ phases (shown in
supplementary \cite{EPAPS}) are stable at 1.0 TPa, and $R\bar{3}m$ is
also stable at 1.8 TPa.  These structures have a group-subgroup
relation which may indicate that oxygen molecules symmetrize upon
increasing pressure near the transformation to polymeric phases.  The
molecular $\zeta$ oxygen is a superconductor at temperatures below 0.6
K \cite{Shimizu1998}.  We have therefore calculated the phonon
linewidth, Eliashberg function $\alpha^2F(\omega)$, electron-phonon
coupling (EPC) constant $\lambda$, and the logarithmic average phonon
frequency $\omega_{log}$, to investigate whether the molecular
$R\bar{3}m$ and polymeric $Cmcm$ phases of oxygen are superconducting.

As shown in Fig.\ \ref{fig:phonon} (upper), the acoustic phonon modes
and the intramolecular LO modes of $R\bar{3}m$ have relatively large
linewidths close to the $\Gamma$ point, while the TO modes have large
linewidths along L--F.  We have used the Allen-Dynes modification of
the McMillan equation for the superconducting transition temperature
\cite{Allen1975},
$T_c=\omega_{log}/1.2\exp[-1.04(1+\lambda)/\lambda-\mu^{\ast}(1+0.62\lambda)]$,
where $\mu^{\ast}$ is the Coulomb pseudopotential, with typical values
of $\mu^{\ast}$ between 0.10 and 0.13. The calculated isotropic EPC
constant for the $R\bar{3}m$ structure at 1.8 TPa is about 0.34 and
$T_c$ varies from 2.1 to 0.6 K, which is close to the value observed
for the $\zeta$ phase of about 0.6 K \cite{Shimizu1998}.

The phonon band structure of $I4_1/acd$ at 2.0 TPa is shown in Fig.\
\ref{fig:phonon} (lower left).  The phonon dispersion along X--P and
P--N is small, indicating that the bonding between the square spirals
is weak.  The phonon dispersion relation of $Cmcm$ at 3.5 TPa (Fig.\
\ref{fig:phonon} lower right) shows the structure to be stable and
highly anisotropic.  Chain-like and layered structures also appear in
other molecular systems, such as CO \cite{Sun2009-CO2,Sun2011-CO} and
N$_2$ \cite{Ma2009-nitrogen}.

We have investigated oxygen under terapascal pressures using the AIRSS
technique and \textit{ab initio} DFT calculations.  Oxygen molecules
persist up to about 1.9 TPa, which is almost one order of magnitude
larger than previously reported \cite{Ma2007} and also much larger
than other diatomic molecules such as H$_2$, N$_2$ and CO.  Oxygen
polymerizes at about 1.9 TPa giving a square spiral-like structure of
space group $I4_1/acd$, which is a semiconductor with a DFT band gap
of about 3.0 eV.  The $R\bar{3}m$ structure is superconducting at 1.8
TPa and has a similar $T_c$ to the $\zeta$ phase.  $R\bar{3}m$ is
slightly more stable than other molecular forms at high pressures and
its metallic oxygen molecules symmetrize before polymerization.  At
about 3.0 TPa, the system collapses to a denser metallic $Cmcm$ phase,
which consists of zig-zag chains.  The stability of molecular and
chain-like phases of highly-compressed solid oxygen demonstrates that
the electron counting rule is effective up to multi-TPa pressures.
Merging of the chains in each layer of $Cmcm$ leads to the formation
of a phase of $Fmmm$ symmetry at 9.5 TPa in which each atom has four
nearest neighbors, which signals the eventual failure of the electron
counting rule to predict the bonding.  The strongly repulsive
interactions between the lone pairs hinders the polymerization of the
molecules and the formation of more highly coordinated structures.
The electronic structure of oxygen changes dramatically with
increasing pressure.  It transforms from the insulating $\epsilon$
phase to the molecular metal $\zeta$ phase, which persists up to 1.9
TPa, where it transforms into the semiconducting $I4_1/acd$ polymeric
phase, eventually returning to metallic behaviour with the appearance
of the $Cmcm$ phase at about 3.0 TPa.

J.S.\ gratefully acknowledges financial support from the Alexander von
Humboldt (AvH) foundation.  C.J.P.\ and R.J.N.\ were supported by the
EPSRC.  The calculations were carried out at {\sc Bovilab@RUB}
(Bochum), on the supercomputers at NRC (Ottawa) and UCL.










\bibliographystyle{prsty}

\end{document}